\documentstyle[12pt,a4]{article}
\begin{document}
\newcommand{\wst}{~{}^{{}^{*}}\llap{$\it w$}}
\newcommand{\wdst}{~~~{}^{{}^{**}}\llap{$\it w$}}
\newcommand{\omegast}{~{}^{{}^{*}}\llap{$\omega$}}
\newcommand{\omegadst}{~~~{}^{{}^{**}}\llap{$\omega$}}
\newcommand{\must}{~{}^{{}^{*}}\llap{$\mu$}}
\newcommand{\mudst}{~{}^{{}^{**}}\llap{$\mu$~}}
\newcommand{\nust}{~{}^{{}^{*}}\llap{$\nu$}}
\newcommand{\nudst}{~{}^{{}^{**}}\llap{$\nu$~}}
\newcommand{\beq}{\begin{equation}}
\newcommand{\eeq}{\end{equation}}
\newcommand{\gfrc}[2]{\mbox{$ {\textstyle{\frac{#1}{#2} }\displaystyle}$}} 
\title {\large Higher Dimensional Static Cosmological Model in Lyra Manifold }
\author { G S Khadekar \thanks{Tel.91-0712-239461,
email:gkhadekar@yahoo.com} and  Avinash R Nagpure \thanks{Department
of Mathematics, Jawarharlal Nehru Arts, Commerce \& Science College,
Wadi (Defence), Nagpur-440 023 (INDIA)}  \\
Department of Mathematics, Nagpur University \\ 
Mahatma Jyotiba Phule Educational Campus, Amravati Road \\ 
Nagpur-440 010 (INDIA)}
\maketitle
\begin{abstract}
A static conformally flat spherically symmetric perfect fluid
cosmological model in Lyra geometry is investigated. 
\end{abstract}
\section{ Introduction}
The geometry proposed by Lyra [1] stands for the Riemannian geometry and
was utilized by Sen [2, 3], Halford [4], Bhamra [5]  and Kalyanshetti
and Waghmode [6] to probe into
some of the aspects of cosmological theory. Having known that the
conformally flat space-time are of particular interest in the context
of petrov classification in view of their degeneracy. In this paper,
we imply the geometry to study the static conformally flat
cosmological model filled with perfect filled.
\subsection{Field Equation}
The field equations in normal gauge for Lyra's manifold as obtained by
Sen [2] are 
\beq
R_{ij}- \frac{1}{2}\; g_{ij} R + \frac{3}{2} \phi_{i} \phi_{j} -
\frac{3}{4} g_{ij} \phi_{k} \phi^{k} = -8 \pi T_{ij}
\eeq
where  $ \phi_{i} $ as a displacement field and the other symbols have
their usual meanings as in Riemannian geometry. We assume the vector
displacement field $ \phi_{i} $ to be the time like constant vector.
\beq
\phi_{i} = (0, 0, 0, 0, \beta)
\eeq
where $ \beta $ is a
constant. Consider the five dimensional  static conformally flat spherically symmetric
perfect fluid represented by the space-time metric
\beq
ds^2 = e^{2 \alpha} \left( dt^2 - dr^2 -r^2 d\theta^2 -r^2 sin^2
\theta d \phi^2 + dm^2 \right)
\eeq
where $ \alpha $ is a function of r alone. The energy momentum tensor
$ T_{ij} $ for perfect fluid distribution is given by 
\beq
T_{ij} = (P + \rho ) V_{i} V_{j} -  P g_{ij}
\eeq
together with 
\beq
g_{ij} V^{i} V^{j}
\eeq
where $ V^{i}$ is the five velocity vector of the fluid and $ P $ and $
\rho $ are the proper pressure and density respectively.\\
By use of equations (2), (4) and (5), the fields equations (1) for the
line element (3) can be written as 
\beq
e^{-2 \alpha}\left( 6 \alpha^2_{1} + \frac{ 6 \alpha_{1}}{r} -
\frac{3}{4} \beta^2 \right) = 8 \pi P 
\eeq
\beq
e^{-2 \alpha}\left(3  \alpha_{11} + 3  \alpha^2_{1} + \frac{ 3 \alpha_{1}}{r} -
\frac{3}{4} \beta^2 \right) = 8 \pi P
\eeq
\beq
e^{-2 \alpha}\left(3  \alpha_{11} + 3  \alpha^2_{1} + \frac{ 6 \alpha_{1}}{r} +
\frac{3}{4} \beta^2 \right) = - 8 \pi \rho
\eeq
where  $ \alpha_{1} = \frac{d \alpha}{dr}, \; \; \alpha_{11} =
\frac{d^2\alpha}{dr^2} $ etc.\\
Eliminating $ P $ between equation (6) and (7) we get
\beq
\alpha_{11} - \alpha_{1}^2 - \frac{\alpha_{1}}{r} = 0
\eeq
which results into the solution
\beq
e^{-\alpha} = ( A r^2 + B )
\eeq
where $ A $ and $ B $ are arbitrary constants. Hence, the static
conformally flat cosmological models in Lyra geometry takes the form 
\beq
ds^2 = ( A r^2 + B )^{-2} \left( dt^2 - dr^2 -r^2 d\theta^2 -r^2 sin^2
\theta d \phi^2 + dm^2 \right)
\eeq
It is interesting to observe that the model (11) is similar to the
static conformally symmetric model obtained by Kalyanshetti and
Waghmode [6] in the Einstein-Cartan theory for a perfect fluid with a
classical description of spin.
\subsection{Some Properties of the Model}
The pressure $ P $ and density $ \rho $ in the model are given by
\beq
8 \pi P = 12 A^2 r^2  -12 A B - \frac{3}{4} \beta^{2} (A r^2 + B)^2
\eeq
\beq
8 \pi \rho = 18 A B - 6 A^2 r^2 - \frac{3}{4} \beta^2 (A r^2 + B)^2
\eeq
The dominant energy conditions,
\beq
\rho = 0,  \; \; \;  \rho - P \ge 0, \; \; \: \rho + P \ge 0. 
\eeq  
For the general perfect fluid distribution leads to 
\beq
5 A B \ge 3 A^2 r^2 , \: \: \; \beta < \frac{4 A }{(A r^2 + B )}
\eeq
If $ A = 0 $ and $ B = 0, $ we have from (15) that   $ \;  \beta^2 < 0 $ so
that the displacement vector $ \phi_{i} $ in Lyra manifold is imaginary
and  thus the model is physically unrealistic. Also when  $ A= 0, $
the  space-time is flat. Also with $ P \ge 0 $ and $ \beta^2 > = 0,$
one obtains
$$ A^2 r^2 \ge A B  \; \: \: $$  so that $$ \frac{5 B}{3 A} > r^2 \ge
\frac{B}{A} $$ The solution is restricted to a comparatively small
region of the space-time.
\subsection{Conclusions}
It is observed that the situation in this case is similar (apart from
a difference of sign) to that discussed by Kalyanshetti and Waghmode
[7] in the Einstein-Cartan theory with number $ \beta^2 $ and therefore,
the displacement  vector $ \phi $ in Lyra manifold plays the role of the
spin density. 
\bibliographystyle{plain}

\end{document}